\documentstyle[psfig,longtable,amssymb]{aipproc}

\renewcommand{\[}{\begin{equation}}
\renewcommand{\]}{\end{equation}}
\newcommand{\intr}{\int d{\bf r} \;}
\def\equ#1{Eq.~(\ref{#1})}
\def\eqs#1#2{Eqs.~(\ref{#1}) and (\ref{#2})}
\begin{document}

\setlength{\unitlength}{1cm}
\noindent \begin{picture}(0,0)
\put(0,2.5){\noindent \sf Presented at the workshop:}
\put(0,2){\noindent \sf ``Fundamental Physics of Ferroelectrics'',}
\put(0,1.5){\noindent \sf Aspen Center for Physics, February 13-20, 2000}
\end{picture}
\title{What makes an insulator different \\ from a metal?}

\author{Raffaele Resta}
\address{INFM -- Dipartimento di Fisica Teorica, \\
Universit\`a di Trieste, Strada Costiera 11, 
I-34014 Trieste, Italy}

%\lefthead{LEFT head}
%\righthead{RIGHT head}
\maketitle

\begin{abstract} The insulating state of matter is characterized by the
excitation spectrum, but also by qualitative features of the electronic
ground state. The insulating ground wavefunction in fact: (i) displays
vanishing dc conductivity; (ii) sustains macroscopic polarization; and (iii)
is {\it localized}. The idea that the insulating state of matter is a
consequence of electron localization was first proposed in 1964 by W. Kohn. I
discuss here a novel definition of electron localization, rather different
from Kohn's, and deeply rooted in the modern theory of polarization. In fact
the present approach links the two features (ii) and (iii) above, by means of
essentially the same formalism. In the special case of an uncorrelated
crystalline solid, the localization of the many--body insulating wavefunction
is measured---according to our definition---by the spread of the Wannier
orbitals; this spread diverges in the metallic limit. In the correlated case,
the novel approach to localization is demonstrated by means of a two--band
Hubbard model in one dimension, undergoing a transition from band insulator
to Mott insulator.  \end{abstract}

\section*{Introduction}

The present contribution, dealing with such a general issue as the difference
between insulators and metals, may appear out of place at a workshop
focussing on the ``Fundamental Physics of Ferroelectrics''. But indeed, the
present results are in a sense the ultimate developments of the theory of
polarization~\cite{modern,rap_a12,Ortiz94,rap100,rap112} based on a Berry
phase~\cite{Berry}, which is crucial to the modern understanding of
ferroelectrics~\cite{William}. The link between the present subject and
polarization is simply stated: insulators sustain nontrivial bulk
polarization, metals do not.

To the present purposes, materials are conveniently divided in only two
classes: insulators and metals. We use here the terms in a loose sense as
synonym of nonconducting and conducting: an insulator is distinguished from a
metal by its vanishing conductivity at low temperature and low frequency. 
This qualitative difference in the dc conductivity must reflect a qualitative
difference in the organization of the electrons in their ground state.  So
the question we are going to address is: is it possible to find a {\it pure
ground state property} which discriminates between an insulator and a metal?
Before proceeding to answer, let me discuss the alternative phenomenological
characterization of the insulating/metallic behavior: instead of making
direct reference to the dc conductivity, we address macroscopic polarization.

Suppose we expose a finite macroscopic sample to an electric field, say
inserting it in a charged capacitor. Then the induced macroscopic
polarization is qualitatively different in metals and insulators. In the
former materials polarization is trivial: universal, material--independent,
due to surface phenomena only (screening by free carriers). Therefore
polarization in metals is {\it not} a bulk phenomenon. The opposite is true
for insulators: macroscopic polarization is a nontrivial,
material--dependent, bulk phenomenon. We can therefore phenomenologically
characterize an insulator, in very general terms, as a material whose ground
wavefunction sustains a bulk macroscopic polarization whenever the
electronic Hamiltonian is non centrosymmetric. From this definition it is
clear that the modern theory of polarization, based on a Berry's phase, can
lead to a better understanding of the insulating state of matter.

This paper is organized as follows. First we briefly outline Kohn's theory of
the insulating state~\cite{Kohn64}. Then we address dipole and localization
for a lone electron in a Born--von--K\`arm\`an periodic box; subsequently, we
apply similar ideas  to an extended system of $N$ electrons in a periodic
box, addressing macroscopic polarization and electronic localization in the
many--body case. We then show how this works for a crystalline system of
independent electrons, and finally---following Ref.~\cite{rap107}---we
demonstrate localization for a model correlated system displaying two
different insulating phases. For the sake of simplicity, we explicitate here
the relevant algebra only for the case of one--dimensional electrons. The
generalization to three--dimensional electrons can be found in
Refs.~\cite{rap_a20,rap_a21,Souza00}.

\section*{The insulating state and Kohn's theory}

Within any classical theory, the electronic responses of insulators and
metals are qualitatively described by ``bound'' and ``free'' charges,
respectively. Microscopic models for such charges are provided by the Lorentz
theory (insulators) and by the Drude theory (metals).  Within the former,
each electron is tied (by an harmonic force) to a particular center; within
the latter, electrons roam freely over macroscopic distances, hindered only
by atomic scattering potentials.  Therefore, from a purely classical model
viewpoint, one explains the insulating/metallic behavior of a material by
means of the localized/delocalized character of the electron distribution.

Switching to quantum mechanics, this clearcut character of the electron
distribution is apparently lost. Textbooks typically explain the
insulating/metallic behavior by means of band structure theory, focussing on
the position of the Fermi level of the given material: either in an band gap
(insulators), or across a band (metals). This picture is obviously correct,
but very limited and somewhat misleading.  

First of all, the band picture applies only to a crystalline material of
independent electrons: a very limited class of insulators indeed. 
Noncrystalline insulators do in fact exist, and the electron--electron
interaction is a fact of nature: in some materials the insulating
behavior is dominated by disorder (Anderson insulators), in some other
materials  the insulating behavior is dominated by electron correlation (Mott
insulators). Further classes of insulators are also known, as {\it e.g.}
excitonic insulators. Therefore, for a large number of insulators, the band
picture is totally inadequate.

Second: even for a material where a band--structure description is adequate,
the simple explanation of the insulating/metallic behavior focusses on the
spectrum of the system, hence on the nature of the low lying electronic {\it
excitations}. Instead, the qualitative difference in the dc conductivity at
low temperature must reflect a qualitative difference in the organization of
the electrons in their {\it ground state}. Such a difference is not evident
in a band--structure picture: the occupied states are of the Bloch form both
in insulators and in metals, and qualitatively rather similar (in particular
those of simple metals and of simple semiconductors).

In a milestone paper published in 1964, Kohn was able to define the
insulating state of matter in a way which in a sense is close to the
classical picture. In fact he gave evidence that electron localization is the
main feature determining the insulating behavior of a many--electron
wavefunction~\cite{Kohn64}, thus restoring the same basic distinction as in
the classical picture: the key is how to define and to measure the degree of
electronic localization, visualizing in a qualitative and quantitative way
the peculiar organization of the electrons which is responsible for the
insulating state of matter.  

As previously stated, a superficial look indicates that electrons are roughly
speaking equally delocalized in insulators and in metals. One needs therefore
a sharp criterion which singles out the relevant character of the
wavefunction. Kohn's criterion is the following~\cite{Kohn64,Kohn68,Souza00}:
the many--electron wavefunction is localized if it breaks up into a sum of
functions $\Psi = \sum_J \Psi_J$ which are localized in essentially
disconnected regions ${\cal R}_J$ of the configuration space. Any two such
$\Psi_J$'s have exponentially small overlap. Under such a localization
hypothesis, Kohn proves that the dc conductivity vanishes.

\section*{Dipole and localization for a single electron}

The dipole moment of any {\it finite} $N$--electron system in its ground
state is a simple and well defined quantity. Given the many--body
wavefunction $\Psi$ and the corresponding single--particle density $n({\bf
r})$ the electronic contribution to the dipole is: \[ \langle {\bf R} \rangle
= \intr {\bf r} \, n({\bf r}) = \langle \Psi | \hat{{\bf R}} | \Psi \rangle ,
\label{dipp} \] where $\hat{{\bf R}} = \sum_{i=1}^N {\bf r}_i$. This looks
very trivial, but we are exploiting here an essential fact: the ground
wavefunction of any bound $N$--electron system is square--integrable and
vanishes exponentially at infinity. Going at the very essence, we simplify
matter at most and we consider in the present Section only a single electron
in one dimension. The dipole (or equivalently the center) of the electronic
distribution is then: \[ \langle x \rangle = \int_{-\infty}^\infty dx \; x \,
|\psi(x)|^2 , \label{dipole} \] where again we understand $\psi(x)$ as a
square--integrable function over ${\mathbb R}$. This is not the way condensed
matter theory works. Because of several good reasons, in either crystalline
or disordered systems it is almost mandatory to assume  BvK boundary
conditions: the wavefunction $\psi(x)$ is periodic over a period $L$, large
with respect to atomic dimensions.

Adopting a given choice for the boundary conditions is tantamount to defining
the Hilbert space where our solutions of Schr\"odinger's equation live.  By
definition, an operator maps any vector of the given Hilbert space into
another vector belonging to the same space: the multiplicative position
operator $x$ is therefore {\it not} a legitimate operator when BvK are
adopted for the state vectors, while any periodic function of $x$ is
legitimate: this is the case {\it e.g.} of the nuclear potential acting on
the electrons.  

\begin{figure}[t]
\centerline{\psfig{file=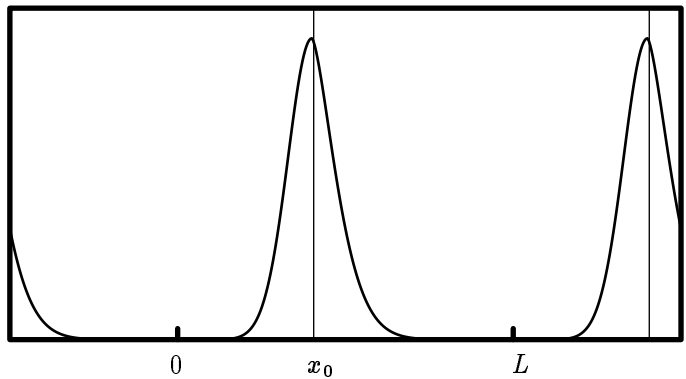,width=9cm}}
\caption{The distribution $|\psi(x)|^2$ of a single--particle orbital
within periodic Born--von--K\`arm\`an boundary conditions}
\label{loc} \end{figure}

Suppose we have an electron distribution such as the one in Fig.~\ref{loc}.
The main issue is then: how do we define the center of the distribution?
Intuitively, the distribution appears to have a ``center'', which however is
defined only modulo the replica periodicity, and furthermore {\it cannot} be
evaluated  simply as in \equ{dipole}, precisely because of BvK.  

Solutions to this and similar problems have been attempted several times:
many incorrect papers---which will not be identified here---have been
published over the years. The good solution has been found by Selloni {\it et
al.} in 1987 by means of a very elegant and far--reaching
formula~\cite{brodo}.  According to them, the key quantity for dealing with
the position operator within BvK is the dimensionless complex number
${\mathfrak z}$, defined as: \[ {\mathfrak z} = \langle \psi  | {\rm
e}^{i\frac{2\pi}{L}x} | \psi \rangle = \int_0^L \! d x \; {\rm
e}^{i\frac{2\pi}{L}x} |\psi(x)|^2 , \label{z} \] whose modulus is no larger
than 1. The most general electron density, such as the one depicted in
Fig.~\ref{loc}, can always be written as a superposition of a function
$n_{\rm loc}(x)$, normalized over $(-\infty, \infty)$, and of its periodic
replicas: \begin{equation} |\psi(x)|^2 = \sum_{m = -\infty}^\infty n_{\rm
loc} (x - x_0 -mL) . \label{replicas} \end{equation} Both $x_0$ and $n_{\rm
loc} (x)$ have a large arbitrariness: we restrict it a little bit by imposing
that $x_0$ is the center of the distribution, in the sense that
$\int_{-\infty}^\infty d x \, x \, n_{\rm loc} (x) =0$. Using
Eq.~(\ref{replicas}), ${\mathfrak z}$ can be expressed in terms of the
Fourier transform of $n_{\rm loc}$ as: \begin{equation} {\mathfrak z} = {\rm
e}^{i\frac{2\pi}{L} x_0} \tilde{n}_{\rm loc} (-\frac{2\pi}{L}) . 
\label{fourier} \end{equation} If the electron is localized in a region of
space much smaller than $L$, its Fourier transform is smooth over reciprocal
distances of the order of $L^{-1}$ and can be expanded as: \begin{equation}
\tilde{n}_{\rm loc} (-\frac{2\pi}{L}) = 1 - \frac{1}{2}
\left(\frac{2\pi}{L}\right)^2 \int_{-\infty}^\infty d x \, x^2 n_{\rm loc}
(x) + {\cal O}(L^{-3}) \label{expansion} . \end{equation} A very natural
definition of the center of a localized periodic distribution $|\psi(x)|^2$
is therefore provided by the phase of ${\mathfrak z}$ as: \begin{equation}
\langle x \rangle = \frac{L}{2\pi} \mbox{Im log}\, {\mathfrak z} ,
\label{center} \end{equation} which is in fact the formula first proposed by
Selloni {\it et al.}~\cite{brodo}. The expectation value $\langle x \rangle$
is defined modulo $L$, as expected since $|\psi(x)|^2$ is BvK periodic.  It
is also worth to observe that for an extremely delocalized state we have
$|\psi(x)|^2 = 1/L$ and ${\mathfrak z} = 0$: hence the center of the
distribution $\langle x \rangle$, according to \equ{center}, is ill--defined,
as one would indeed expect.

So far, we have not specified which Hamiltonian we were addressing when
discussing electron distributions $|\psi(x)|^2$ of the kind depicted in
Fig.~\ref{loc}. It is however obvious to imagine that the wavefunction
$\psi(x)$ is the eigenstate of a (periodically repeated) potential well of
suitable shape. Suppose for a moment we are {\it not} adopting BvK boundary
conditions, having thus only a genuinely isolated potential well. In this
case the eigenstates can belong to two different classes: bound (localized)
states, and scattering (delocalized) states. The distinction is a
qualitatively clearcut one, and can be stated in several ways. One of them is
to consider the second cumulant moment, or spread: \[ \langle x^2
\rangle_{\rm c} =  \langle x^2 \rangle - \langle x \rangle^2 =
\int_{-\infty}^\infty dx \; x^2 \, |\psi(x)|^2 - \left( \int_{-\infty}^\infty
dx \; x \, |\psi(x)|^2 \right)^2 , \label{cum2} \] which is finite for bound
states and divergent (when using appropriate normalizations) for scattering
ones.  But if we study the same potential well within BvK, the qualitative
distinction is lost: all states appear in a sense as ``delocalized'' since
all wavefunctions $\psi(x)$ are periodic over the BvK period. And in fact the
integrals in \equ{cum2} become ill defined.

The main issues therefore are: How do we distinguish between localized and
delocalized states within BvK? In case of a localized state, how we actually
measure the amount of localization? In the literature, such issues have been
previously addressed by means of the participation
ratio~\cite{participation}.  The complex number ${\mathfrak z}$, whose phase
provides the center of the distribution, \equ{center}, is our key
to addressing localization: it is enough to consider its modulus. It has
already been observed that $|{\mathfrak z}|$ is bounded between 0 and 1, and
that $|{\mathfrak z}|$ equals zero for an extremely delocalized state with
$|\psi(x)|^2 = 1/L$. If we take instead an extremely localized state, with
$n_{\rm loc} (x) = \delta(x)$, it is straightforward to get $|{\mathfrak z}|
= 1$. It is therefore natural to measure localization by means of the
negative of the logarithm of $|{\mathfrak z}|$: it is a nonnegative number,
equal to zero in the case of extreme localization, and divergent in the case
of extreme delocalization. A glance at \equ{expansion} yields: \[ \log
|{\mathfrak z}| \simeq  - \frac{1}{2} \left(\frac{2\pi}{L}\right)^2
\int_{-\infty}^\infty d x \, x^2 n_{\rm loc} (x) , \] hence a natural
expression for measuring the actual spread within BvK is: \[ \langle x^2
\rangle_{\rm c} =  \langle x^2 \rangle - \langle x \rangle^2 = -
\left(\frac{L}{2\pi}\right)^2 \log |{\mathfrak z}|^2 . \label{spread} \]
Having in mind again the eigenstates of a potential well, we can study the
expression in \equ{spread} as a function of $L$. For a localized state, the
shape of of $n_{\rm loc} (x)$ can be taken as $L$--independent for large $L$,
hence \equ{spread} goes to a finite limit, which is the ``natural'' spread of
the distribution. Quite on the contrary, for a delocalized state the
distribution is smeared all over the $(0,L)$ segment, preserving the norm
over one period: therefore ${\mathfrak z}$ goes to zero and the spread
diverges in the large--$L$ limit.

\section*{Dipole and localization for many electrons}

So much about the one--electron problem: we are now going to consider a
finite density of electrons in the periodic box. To start with, irrelevant
spin variables will be neglected, and a system of spinless electrons in one
dimension is considered. Even for a system of independent electrons, our
approach takes a simple and compact form if a many--body formulation is
adopted; BvK imposes periodicity in each electronic variable separately. Our
interest is in studying a bulk system: $N$ electrons in a segment of length
$L$, where eventually the thermodynamic limit is taken: $L \rightarrow
\infty$, $N \rightarrow \infty$, and $N/L = n_0$ constant.

We start defining the one--dimensional analogue of $\hat{\bf R}$ of
\equ{dipp}, namely, the multiplicative operator $\hat{X} = \sum_{i=1}^N x_i$,
and the complex number \begin{equation} {\mathfrak z}_N = \langle \Psi | {\rm
e}^{i\frac{2\pi}{L} \hat{X}} | \Psi \rangle .  \label{general} \end{equation}
It is obvious that the operator $\hat{X}$ is ill--defined in our Hilbert
space, while its complex exponential appearing in \equ{general} is well
defined. The main result of Ref.~\cite{rap100} is that the ground--state
expectation value of the position operator is given by the analogue of
\equ{center}, namely: \begin{equation} \langle X \rangle = \frac{L}{2\pi}
\mbox{Im ln } {\mathfrak z}_N  , \label{main} \end{equation} a quantity
defined modulo $L$ as above.

The right--hand side of Eq.~(\ref{main}) is not simply the expectation value
of an operator: it is the {\it phase} of it, converted into length units by
the factor $L/(2\pi)$. This phase can be called a single--point
Berry phase, for reasons explained elsewhere~\cite{rap101,rap_a20,rap_a21}.
Furthermore, the main ingredient of \equ{general} is the expectation value
of the multiplicative operator ${\rm e}^{i\frac{2\pi}{L} \hat{X}}$: it is
important to realize that this is a genuine {\it many--body} operator. In
general, one defines an operator to be one--body whenever it is the {\it sum}
of $N$ identical operators, acting on each electronic coordinate separately:
for instance, the $\hat{X}$ operator is such. In order to express the
expectation value of a one--body operator the full many--body wavefunction is
not needed: knowledge of the one--body reduced density matrix $\rho$ is
enough: I stress that, instead, the expectation value of ${\rm
e}^{i\frac{2\pi}{L} \hat{X}}$ over a correlated wavefunction {\it cannot} be
expressed in terms of $\rho$, and knowledge of the $N$-electron wavefunction
is explicitly needed. In the special case of a single--determinant, the
$N$-particle wavefunction is uniquely determined by the one--body reduced
density matrix $\rho$ (which is the projector over the set of the occupied
single--particle orbitals): therefore the expectation value $\langle X
\rangle$, \equ{main}, is uniquely determined by $\rho$. But this is peculiar
to uncorrelated wavefunctions only.

The expectation value $\langle X \rangle$ is extensive, as the dipole in
\equ{dipp}. For the corresponding intensive quantity we borrow from
Ref.~\cite{Souza00} a useful notation: \[ \langle x \rangle_{\rm c} = \langle
X \rangle / N = \frac{L}{2\pi N} \mbox{Im ln } {\mathfrak z}_N , \label{cum1}
\] where the subscript means ``cumulant''. The quantity $\langle x
\rangle_{\rm c}$ goes to a well defined termodynamic limit, which is in fact
proportional to the macroscopic polarization of the system.  This result is
proved in Ref.~\cite{rap107}; its three dimensional generalization is
discussed in Refs.~\cite{rap_a20,rap_a21,Souza00}.

We stress that nowhere have we assumed crystalline periodicity. Therefore our
definition of $\langle x \rangle_{\rm c}$ is very general: it applies to any
condensed system, either ordered or disordered, either independent--electron
or correlated. In the special case of a crystalline system, either
interacting or noninteracting, the present approach can be shown equivalent
to the previous formulations of polarization
theory~\cite{modern,rap_a12,Ortiz94,rap100,rap_a20,rap_a21,Souza00}.

We are now ready to discuss electron localization in a condensed system: the
present view is the one of Refs.~\cite{rap107,rap_a21}, recently reexamined
by Souza {\it et al.}~\cite{Souza00}, who also discuss its relationship to
Kohn's localization\cite{Kohn64}. This view is based on the modulus of
${\mathfrak z}_N$, in full analogy with the previous Section about the single
electron.  We define therefore an intensive quantity, the second cumulant
moment, by analogy with \eqs{spread}{cum1}: \[ \langle x^2 \rangle_{\rm c} =
- \frac{1}{N} \left(\frac{L}{2\pi}\right)^2 \log |{\mathfrak z}_N|^2 ,
\label{guess} \] where again the notation is borrowed from
Ref.~\cite{Souza00}.  This second moment is a very meaningful measure of
electron localization in the electronic ground wavefunction, and enjoys two
important properties: (1) When applied to a crystalline system of independent
electrons, we recover an important gauge--invariant quantity which controls
the Marzari--Vanderbilt localization~\cite{Marzari97}; (2) Even for more
general systems, correlated and/or disordered, $\langle x^2 \rangle_{\rm c}$
assumes a finite value in insulators, and diverges in metals. Indeed in a
metal the modulus of ${\mathfrak z}_N$ goes to zero in such a way that its
phase is ill defined, and hence macroscopic polarization is ill defined as
well. We have emphasized throughout this paper that one of the main
phenomenological features differentiating insulators from metals is that the
former materials sustain a nontrivial bulk polarization, while the latter do
not. The complex number ${\mathfrak z}_N$ provides the key formal link
between polarization and localization, via its phase and its modulus.

\section*{Noninteracting electrons}

\begin{figure}
\begin{center}
       \setlength{\unitlength}{1cm}
       \begin{picture}(14,3.5) % figure dimensions
\put(0,0){\includegraphics{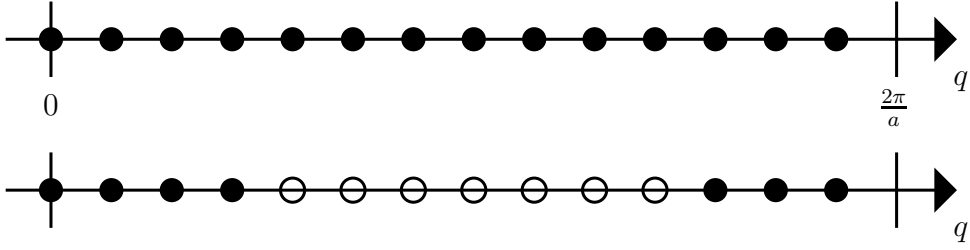}}
\put(0.9,2){$0$}
\put(12,2){$\frac{2\pi}{a}$}
\put(13,2.4){$q$}
\put(13,0.4){$q$}
       \end{picture}
\caption{Discrete $q$ vectors in the reciprocal cell in one dimension, where
a BvK periodicity of $M=14$ crystal cells has been chosen. Black circles
correspond to occupied $q$'s, and empty circles to unoccupied ones. Top:
insulator, with $N=2M$ (filled band). Bottom: metal, with  $N=M$
(half--filled band).
\label{fig:discre2}}
\end{center}
\end{figure}

For independent electrons, we may write the many--body wavefunction $\Psi$ as
a Slater determinant of Bloch orbitals, but in the case of a metal not all
the Bloch vectors in the reciprocal cell correspond to occupied orbitals:
this fact is of overwhelming importance. We consider the simple case of one
band in one dimension, whose Bloch vectors are illustrated in
Fig.~\ref{fig:discre2}, imposing BvK boundary conditions over $M$ cristal
cells: $L = Ma$, where $a$ is the lattice constant.  We restore electron spin
here: we get an insulator if the number of electrons $N$ equals $2M$ (filled
band, top sketch), and a metal if $N=M$ (half--filled band, bottom sketch). 
Both in the insulating and in the metallic case the $N$--electron
wavefunction is a Slater determinant of size $N$, built of $N/2$ doubly
occupied spatial orbitals. Following the same algebra as in
Refs.~\cite{rap100,rap_a20,rap_a21}, the complex number ${\mathfrak z}_N$ can
be written in any case (insulator or metal) by means of the determinant of a
matrix \[ {\mathfrak z}_N = ( \mbox{det} \; {\cal S} )^2, \label{det1} \]
whose elements are \[ {\cal S}_{q_s,q_{s'}} = \frac{1}{a} \int dx \;
\psi^*_{q_{s}}(x) \psi_{q_{s'}}(x) {\rm e}^{i \frac{2\pi}{Ma}x} ,
\label{number} \] and these elements are nonzero whenever $s=s'+1$.  

In the insulating case, owing to complete filling, both $s$ and $s'$ run over
all the $M$ values: for any given $s$, there is always one (and only one)
$s'$ such that the matrix element in \equ{number} is nonzero. That means that
in any row of the ${\cal S}$ matrix---whose size is $M\!\times\!M$---there is
one, and only one, nonvanishing element. Under these circumstances, the
determinant factors as a product of $M$ numbers: \[ \mbox{det} \; {\cal S} =
\prod_{s = 0}^{M-1} \frac{1}{a} \int dx \; \psi^*_{q_{s+1}}(x)  \psi_{q_s}(x)
{\rm e}^{i \frac{2\pi}{Ma}x} , \] where the identity $\psi_{q_M}(x) \equiv
\psi_{q_0}(x)$ is understood (periodic gauge). All the factors are
nonvanishing, and the logarithm of ${\mathfrak z}_N$ is therefore a finite
number. It can be shown that the $N \rightarrow \infty$ limit of $\langle x^2
\rangle_{\rm c}$ coincides with the spread of the optimally localized Wannier
functions, as defined by Marzari and Vanderbilt~\cite{Marzari97}.

The metallic case is very different. Since not all the $q_s$ vectors are
occupied, the indices $s$ and $s'$ run over a subset of the $M$ values
(Fig.~\ref{fig:discre2}): the matrix  ${\cal S}$ is of size
$M/2\!\times\!M/2$. There is one of the two $q_s$ at the Fermi level, for
which the integrals in \equ{number} are all vanishing, for any {\it occupied}
$s'$. Therefore the matrix ${\cal S}$ has a row of zeros, and its determinant
vanishes: its phase, and hence macroscopic polarization, is ill defined.
The logarithm of ${\mathfrak z}_N$ is formally $-\infty$, and the
spread $\langle x^2 \rangle_{\rm c}$  diverges to $+\infty$. This is what we
expected for a metal; the nontrivial fact is that it diverges {\it even at
finite $N$}, while on general grounds we only expected it to diverge in the
thermodynamic ($N \rightarrow \infty$) limit.

So far, the compact and elegant expression of \equ{guess} has been proved to
be appropriate to discriminate between insulators and metals only for a
crystalline systems of independent electrons. For the---much more
interesting---general case of a correlated and/or disordered system, we {\it
postulate} that \equ{guess} performs the same task. The postulate is based on
the general argument---much stressed above---about macroscopic polarization:
well defined in insulators, ill defined in metals (as a bulk property). The
correctness of this postulate has been verified by Resta and Sorella
\cite{rap107} for a one--dimensional model of a correlated crystal. Work on a
model disordered system is in progress.  Other very interesting discussions
about the physical meaning of $\langle x^2 \rangle_{\rm c}$ and its
relationships to the insulating/metallic character of the system can be found
in a paper of Souza {\it et al.} \cite{Souza00}.

\section*{Localization in a model correlated system}

We review here the very recent work of Resta and Sorella \cite{rap107}, where
\equ{guess} is implemented for a one--dimensional two--band Hubbard model at
half filling, intended to mimic an insulator having a mixed ionic/covalent
character, and whose ground wavefunction is explicitly correlated.  The
macroscopic polarization of this model system was studied in
Refs.~\cite{rap87,Ortiz95}.  To the present purpose, it is enough to study the
centrosymmetric geometry: polarization is zero, the wavefunction is real, and
the phase of ${\mathfrak z}_N$ is either 0 or $\pi$. 

The model has a very interesting behavior as a function of $U$: at small $U$
it is a band insulator, while at a critical $U_c$ undergoes a transition to a
Mott--like insulating phase. In the centrosymmetric geometry, ${\mathfrak
z}_N$ is a real number, which changes sign at $U_c$. Its phase $\gamma$ ({\it
i.e.} the single--point Berry phase~\cite{rap101,rap_a20,rap_a21}) jumps
therefore by $\pi$: it turns out that this occurrence is the main fact
signalling the transition: the topological quantum number $\gamma/\pi$ can be
used as an order parameter to identify the two different phases of the
system~\cite{Thouless,Aligia}.  At low values of $U$ our model is a band
insulator, while above $U_c$ is a Mott--like insulator.  What happens to the
second cumulant moment (squared localization length) $\langle x^2
\rangle_{\rm c}$ as a function of $U$?

\begin{figure} \centerline{\psfig{file=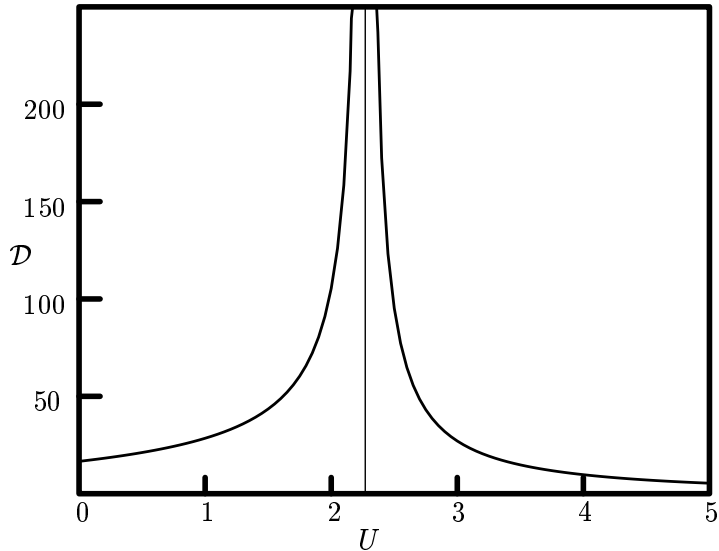,width=9cm}}
\caption[1]{Dimensionless localization parameter,
Eq.~(\protect\ref{dimensionless}), as a function of $U$, after
Ref.~\protect\cite{rap107}, where the effective value of $N$ is
800.  The divergence of the
localization length at the Mott transition is perspicuous.} \label{varu}
\end{figure}

The results are shown in Fig.~\ref{varu} in terms of the dimensionless
quantity \[ {\cal D}_N = - N \log |{\mathfrak z}_N|^2 , \label{dimensionless}
\] such that the squared localization length, \equ{guess} is; \[  \langle x^2
\rangle_{\rm c} = \frac{1}{(2\pi n_0)^2} \lim_{N \rightarrow \infty} {\cal
D}_N .\]  At $U=0$ the system is noninteracting, and the squared localization
length $\langle x^2 \rangle_{\rm c}$ coincides with the second moment of the
(optimally localized) Wannier function of the occupied band.

In the correlated case at $U \neq 0$ no Wannier analysis can be performed;
yet $\langle x^2 \rangle_{\rm c}$ mantains its role as a meaningful measure
of the localization of the electronic wavefunction as a whole. The
localization length increases with $U$ below the transition, diverges at the
transition point $U_c$, and becomes localized again in the highly correlated
regime, where in fact $\langle x^2 \rangle_{\rm c}$ decreases with increasing
$U$. Since the localization length remains finite at all values of $U$
different from $U_c$, the system is always insulating except at the
transition point; the two insulating phases are topologically different and
correspond to a qualitatively different organization of the electrons in the
wavefunction~\cite{Ortiz95}.

What about the transition point? According to the previously stated
viewpoint, the delocalized behavior implies a metallic character of the
many--electron system. Notice that $\langle x^2 \rangle_{\rm c}$ is a pure
ground state property and apparently carries no information about the
excitation spectrum of the system. Yet we have explicitly verified---by
exploiting the metastability of the Lanczos algorithm---that at the critical
$U$ value there is indeed a level crossing. At the transition value $U_c$ the
ground state is twice degenerate and the lowest lying excitation (at constant
$N$) has vanishing energy.
\section*{Acknowledgments}

Discussions with R. M. Martin, F. Mauri, Q.  Niu, G. Ort\'{\i}z, A.
Pasquarello, S. Sorella, I. Souza, and D. Vanderbilt are gratefully
acknowledged. Part of this work was performed at the 1998 workshop ``Physics
of Insulators'' at the Aspen Center for Physics. Partly supported by the
Office of Naval Research, through grant N00014-96-1-0689.

\end{document}